\documentclass[prc,aps,twocolumn,preprintnumbers,amsmath,amssymb,floatfix,superscriptaddress]{revtex4}
\usepackage[usenames]{color}
\usepackage{amsmath}
\usepackage{graphicx}
\usepackage{float,epsfig}
\usepackage{lscape}
\newcommand\be{\begin{equation}}
\newcommand\ee{\end{equation}}
\newcommand\bea{\begin{eqnarray}}
\newcommand\eea{\end{eqnarray}}

\usepackage[usenames]{color}
\usepackage{ulem}

\begin{document}

\title{From dilute matter to the equilibrium point in the
energy--density--functional theory}
\author{C.J. Yang}
\affiliation{Institut de Physique Nucl\'eaire, IN2P3-CNRS, Universit\'e Paris-Sud,
F-91406 Orsay Cedex, France}
\author{M. Grasso}
\affiliation{Institut de Physique Nucl\'eaire, IN2P3-CNRS, Universit\'e Paris-Sud,
F-91406 Orsay Cedex, France}
\author{D. Lacroix}
\affiliation{Institut de Physique Nucl\'eaire, IN2P3-CNRS, Universit\'e Paris-Sud,
F-91406 Orsay Cedex, France}

\begin{abstract}
Due to the large value of the scattering length in nuclear systems, standard
density--functional theories based on effective interactions usually fail to
reproduce the nuclear Fermi liquid behavior both at very low densities and
close to equilibrium. Guided on one side by the success of the Skyrme
density functional and, on the other side, by resummation techniques used in
Effective Field Theories for systems with large scattering lengths, a new
energy--density functional is proposed. This functional, adjusted on
microscopic calculations, reproduces the nuclear equations of state of
neutron and symmetric matter at various densities. Furthermore, it provides
reasonable saturation properties as well as an appropriate density
dependence for the symmetry energy.
\end{abstract}

\maketitle

Various properties of nuclear matter are intimately related to nuclear
phenomena. Such link strongly guides us in constraining effective
interactions within the energy--density functional (EDF) theory, especially
close to the equilibrium density of symmetric matter $\rho_0$, which roughly
corresponds to central densities of medium--mass and heavy nuclei.
Reproducing simultaneously the equation of state (EOS) of symmetric and pure
neutron matter is an important step for producing interactions tailored to
treat both stable and neutron--rich unstable nuclei or even, in the most
extreme cases, the isospin--asymmetric systems located in the crust of
neutron stars. In phenomenological EDFs, attention is usually not paid to
correctly describe the very low--density regime and only density scales of
interest in nuclear phenomena are explored.

It is known that very dilute neutron matter can be described as an
expansion in $k_{N}a$ given by Lee and Yang in Ref. \cite{lee}, where $k_N$ is the neutron
Fermi momentum and $a$ the neutron-neutron $^1S_0$ scattering length, equal
to -18.9 fm. Such low--density regime is well described, by construction, in
all \textit{ab--initio} EOSs derived within effective field theories (EFTs) 
\cite{hammer,schafer}. To our knowledge, it is however never reproduced by
phenomenological EOSs directly adjusted around the saturation point, such
as, for instance, Skyrme \cite{skyrme,vauth} or Gogny \cite{gogny1,gogny2}
EOSs, even in those cases where a special care is taken in well describing
the EOS of neutron matter \cite{chaba,d1n}.

The first two terms of Lee-Yang expansion are 
\begin{equation}
\frac{E_{NM}}{N}=\frac{\hbar ^{2}k_{N}^{2}}{2m}\left[ \frac{3}{5}+\frac{2}{3\pi }%
(k_{N}a)+\frac{4}{35\pi ^{2}}(11-2\ln 2)(k_{N}a)^{2}\right] ,  \label{nm}
\end{equation}%
where $m$ and $N$ are the nucleon mass and the number of neutrons,
respectively, and the Fermi momentum is related to the density $\rho $ by $%
k_{N}=(3\pi ^{2}\rho )^{1/3}$. However, this expression is valid only in the
very dilute regime, that is for $|ak_{N}|<<1$. In the nuclear case, owing to
the large value of the scattering length, this limit corresponds to
extremely low densities, definitely far from typical nuclear densities.

Recently, efforts were made to connect phenomenological EDFs and EFTs based
on contact interactions \cite{Fur12}. Similarly to what is done in the
many--body Dyson expansion \cite{FW}, the total energy computed with EDF
theories can be written as an expansion 
\begin{equation}
E(k_{N})=E^{(1)}(k_{N})+E^{(2)}(k_{N})+\cdots , \label{eq:exp}
\end{equation}%
where the leading order $E^{(1)}$ is the mean--field-EDF (MF--EDF) energy and $E^{(2)}
$ is the second--order perturbation energy treated with proper regularization. This
approach was recently applied to obtain beyond--mean--field functionals
reproducing nuclear matter properties \cite{mog,new}. Within the EDF theory,
we discuss here the possibility of combining an explicit $\rho $ dependence
in an effective interaction with the constraint of correctly reproducing the
very low--density regime, that is, of reproducing the first two terms of the
Lee-Yang expansion. To make this, we first analyze symmetric and neutron
matter with a simplified Skyrme interaction which contains the minimal terms
for reproducing the saturation point at the leading order of the Dyson
equation (MF approximation), 
\begin{equation}
v(\vec{r})=t_{0}(1+x_{0}P_{\sigma })\delta (\vec{r})+\frac{1}{6}%
t_{3}(1+x_{3}P_{\sigma })\rho ^{\alpha }\delta (\vec{r}),
\label{sk}
\end{equation}%
where $P_{\sigma }=\frac{1}{2}(1+\sigma _{1}\cdot \sigma _{2})$ is the
spin--exchange operator, and $t_{0}$, $t_{3}$, $x_{0}$, $x_{3}$, and $\alpha 
$ are parameters. 
\begin{table}[tbp]
\begin{tabular}{cccc}
\hline
$t_0$ (MeV fm$^3$) & $t_3$ (MeV fm$^4$) & $x_0$ & $x_3$ \\ \hline\hline
-1803.93 & 12911.00 & -4.46 & -139.40 \\ \hline
\end{tabular}%
\caption{Values of fitted parameters $x_{0}$ and $x_{3}$ and corresponding $%
t_{0}$ and $t_{3}$ values.}
\label{tab-0}
\end{table}

The interaction given by Eq. (\ref{sk}) generates at each order of the Dyson
equation a given density functional. From the MF density functional, one can
deduce the MF--EOS for neutron matter, that we report in terms of the Fermi
momentum $k_N$, 
\begin{eqnarray}
\frac{E_{NM}^{(1)}}{N}&=&\frac{3}{10}\frac{\hbar^2}{m}k_N^2+\frac{1}{12\pi^2}%
t_0(1-x_0)k_N^3  \notag \\
&+&\frac{1}{24}t_3 (1-x_3)\left(\frac{1}{3\pi^2}\right)^{\alpha+1}k_N^{3%
\alpha+3}.  \label{meanfieldnm}
\end{eqnarray}
The term in $k_N^3$ in Eq. (\ref{nm}) has the same $k_N$ dependence as the
contribution generated at leading order by the $t_0$ term of the
interaction, with the relation 
\begin{equation}
t_0(1-x_0)=4\pi \hbar^2a/m.  \label{t0}
\end{equation}
It is indeed possible to mimic also the following term ($k_N^4$) of the
Lee-Yang expansion still using the density functional provided by the MF.
One possibility would be to use the non--local interaction proposed in Ref. 
\cite{gez}. We follow here another direction and use the explicit density
dependence of the effective interaction. The choice $\alpha=1/3$
leads to a $k_N^4$ term in the EOS (such direction was already explored for
instance in Refs. \cite{bul1,bul2} in the case of dilute Fermi gases in the
unitary regime). For $\alpha=1/3$, by comparing Eqs. (\ref{nm}) and (\ref%
{meanfieldnm}), it must hold 
\begin{equation}
t_3(1-x_3)=\frac{\hbar^2}{m}\frac{144}{35}(3\pi^2)^{1/3}(11-2 \ln 2)a^2.
\label{t3}
\end{equation}
Eqs. (\ref{t0}) and (\ref{t3}) leave in this case only two free
parameters in Eq. (\ref{sk}). 

We write now the MF--EOS for symmetric matter in terms of the Fermi momentum 
$k_F$ ($k_F=(3 \pi^2 \rho /2 )^{1/3}$), 
\begin{equation}
\frac{E^{(1)}_{SM}}{A}=\frac{3}{10}\frac{\hbar^2}{m}k_F^2+\frac{1}{4}\frac{%
t_0}{\pi^2}k_F^3+\frac{1}{16}t_3 \left(\frac{2}{3\pi^2}\right)^{%
\alpha+1}k_F^{3\alpha+3},  \label{meanfieldsm}
\end{equation}
where $A$ indicates the number of nucleons. We adjust the remaining two free
parameters of the interaction to reproduce a benchmark EOS for symmetric
matter, that we have chosen as the SLy5--MF \cite{chaba} EOS. This fit was successfully
performed leading to the parameters listed in Table I. The incompressibility
modulus associated to such EOS is 240.52 MeV. This adjustment is however a
partial success. It indeed allows us to reproduce by construction the
correct (very) low--density behavior in neutron matter and a reasonable
saturation point in symmetric matter but, unfortunately, the resulting
neutron matter EOS is completely wrong beyond the low--density limit.
To obtain a satisfactory EOS for neutron matter at higher densities, the 
scattering length should be treated as a free parameter. However, by adjusting on 
a benchmark EOS for neutron matter, such fit would provide values of $a$ very far 
from -18.9 fm as noted in Ref. \cite{Fur12}. 

Along the line of the expansion given in Eq. (\ref{eq:exp}), we then 
enrich the functional by adding to the MF EOSs the corresponding
second--order contributions, calculated in Refs. \cite{mog,new}, by keeping
only the finite parts (after dimensional regularization). For neutron matter, 
the second--order contribution is  
\begin{figure}[tbp]
\includegraphics[scale=0.32]{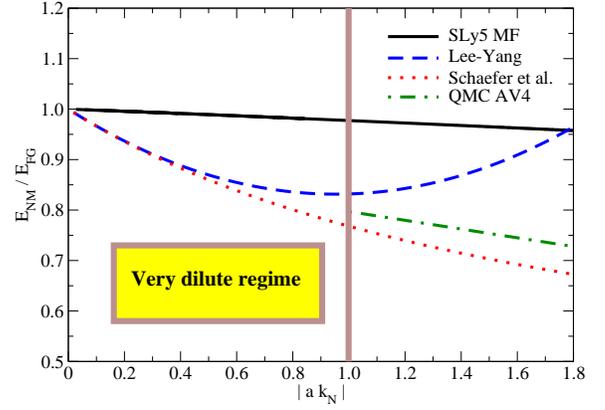}
\caption{(Color online) Neutron matter energy divided by the free 
gas energy $E_{FG}$, obtained with the first two terms of Lee-Yang formula, Eq. (%
\protect\ref{nm}), (blue dashed line), a resumed formula of Ref. 
\protect\cite{schafer} (red dotted line), the SLy5--mean--field EOS (black solid
line), and the QMC AV4 calculations of Ref. \protect\cite{geze} (green
dot--dashed line).}
\label{leeyang}
\end{figure}

\begin{equation}
\frac{E_{NM}^{(2)}}{N}=c[t_{0}^{2}(1-x_{0})^{2}k_{N}^{4}+f_{03}k_{N}^{3%
\alpha +4}+f_{3}k_{N}^{6\alpha +4}],  \label{secordernm}
\end{equation}%
where $c=m^*(11-2\ln 2)/(280\pi ^{4}\hbar ^{2})$ 
and $m^*$ is the effective mass.
Expressions for the coefficients $f_{03}$ and $f_{3}$ may be found in Ref. \cite{new}%
.
The second--order correction for neutron matter contains a $k_{N}^{4}$
contribution which is provided by the $t_{0}^{2}(1-x_{0})^{2}$ term. Using
the constraint of Eq. (\ref{t0}), one recovers 
the second term of the Lee--Yang expression, and could thus guarantee a
correct low--density behavior. It is also interesting to observe that, to
guarantee that the smallest $k_{N}$ dependences in the EOS are $k_{N}^{3}$
and $k_{N}^{4}$, as in the Lee-Yang expansion, the parameter $\alpha $
should be larger than 1/3, which is not always the case for Skyrme
forces. By using the second--order EOS for neutron matter, we
have performed the adjustment of the Skyrme parameters on a benchmark EOS, 
keeping the additional constraint of reproducing the first two terms of the
Lee-Yang expansion in the case of neutron matter. 
In spite of the correct $k_{N}$ dependence provided by the second--order
contribution, it turns out that such fit is not 
successful, 
 except in the case
where the value of $a$ is kept free. However, this leads to an adjusted
value of $a$ close to -1 fm, definitely very far from the physical one. This
direction was then rejected.
\begin{figure}[tbp]
\includegraphics[scale=1.0]{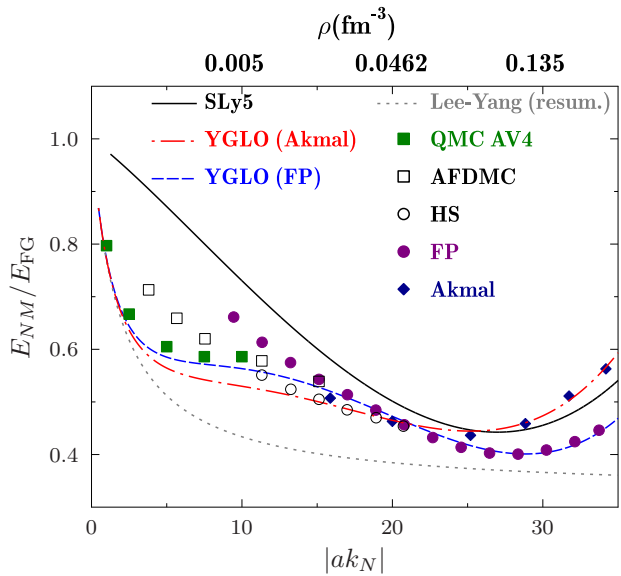}
\caption{(Color online) Energy of neutron matter divided by the free gas
energy $E_{\mathrm{FG}}$ obtained with the two fits of this work, YGLO (FP)
(blue dashed line) and YGLO (Akmal) (red dot--dashed line). The SLy5--MF EOS
(black solid line) is also plotted together with the QMC AV4 points of Ref. 
\protect\cite{geze} (green squares), the Friedman-Pandharipande (FP) results
of Ref. \protect\cite{panda} (violet circles), the Akmal et al. results of
Ref. \protect\cite{akmal} (blue diamonds). The Auxiliary--Field results
(AFDMC) of Ref. \protect\cite{Gan08} and the N2LO calculation of Ref. 
\protect\cite{Heb10} (HS) are also shown, respectively, with black open squares
and black open circles. As an indication, 3 values of $\protect\rho $ are
provided in the upper horizontal axis.}
\label{neutronm}
\end{figure}

The first terms of the Lee-Yang expansion provide a correct behavior for neutron 
matter only for $|a k_N | << 1$. 
To produce expressions that are meaningful also at typical nuclear density
scales, still keeping the good property of correctly reproducing the
low--density regime, various resumed expressions  have been proposed in EFT \cite{steele,schafer,kaiser1}.
In its simplest form, a resumed formula may be for instance written as \cite%
{schafer} 
\begin{equation}
\frac{E_{NM}}{N}=\frac{\hbar ^{2}k_{N}^{2}}{2m}\left[ \frac{3}{5}+\frac{2}{3\pi }%
\frac{k_{N}a}{1-6k_{N}a(11-2\ln 2)/(35\pi )}\right] .  \notag  \label{res}
\end{equation}%
In Fig. \ref{leeyang}, the energy obtained with this expression is compared
to Eq. (\ref{nm}) as well as to recent Quantum Monte Carlo (QMC)
calculations \cite{geze} based on realistic nuclear forces \cite{wiringa}.
The different curves are in agreement at very low densities. As an
illustration, the Skyrme SLy5 MF EOS is also shown. It completely fails to reproduce the low density behavior.
On the other side, 
Skyrme EDFs are recognized for reproducing remarkably well the equilibrium 
features of symmetric matter and the incompressibility modulus. This is in particular due to the explicit two--body 
density--dependent term that is introduced in their expression. 

Guided by the fact that the second--order $t_{0}$ contribution leads to the
correct $k_{N}$ dependence in neutron matter (which can be associated to the
second term of the Lee-Yang expansion), guided on one side by the resumed
formulae of Refs. \cite{steele,schafer,kaiser1}, and on the other side by
the good properties of velocity--dependent and density--dependent terms in
Skyrme forces, we propose a local density functional that includes
resummation to all orders in an effective way. We write the energy as $E=\int
\{\mathcal{K}({\mathbf{r}})+\mathcal{V}({\mathbf{r}})\}d^{3}{\mathbf{r}}$,
where $\mathcal{K}$ is the kinetic term. The functional ${\cal V}$, 
to be used for symmetric ($\beta =1$)
and neutron ($\beta =0$) matter, is given by
\begin{equation}
\mathcal{V}=\frac{B_{\beta }\rho ^{2}}{1-R_{\beta }\rho ^{1/3}+C_{\beta
}\rho ^{2/3}}+D_{\beta }\rho ^{8/3}+F_{\beta }\rho ^{\alpha +2}.
\end{equation}%
This functional leads to the following EOS, 
\begin{equation*}
\frac{E}{A}=K_{\beta }+\frac{B_{\beta }\rho }{1-R_{\beta }\rho
^{1/3}+C_{\beta }\rho ^{2/3}}+D_{\beta }\rho ^{5/3}+F_{\beta
}\rho ^{\alpha +1},
\end{equation*}%
where $A$ becomes $N$ in the case of neutron matter, and $K$ is
the kinetic contribution. We denote such functional with the acronym YGLO 
(for ``Yang-Grasso-Lacroix-Orsay''). The two parameters $B_{\beta }$ and $%
R_{\beta }$ are fixed by imposing to recover the Lee-Yang formula at low
density (the analog of Eq. (\ref{nm}) for symmetric matter may be found in
Refs. \cite{FW,Bis73}). This gives the constraints 
\begin{equation*}
B_{\beta }=2\pi \frac{\hbar ^{2}}{m}\frac{(\nu -1)}{\nu }a,~R_{\beta }=\frac{%
6}{35\pi }\left( \frac{6\pi ^{2}}{\nu }\right) ^{\frac{1}{3}}(11-2\ln 2)a
\end{equation*}%
where $\nu =2$ (4) is the degeneracy for $\beta =$ 0 (1), and $a$ is the corresponding scattering length. In principle, the scattering length used for
symmetric matter should be an average over all the channels. Following the
discussion of Ref. \cite{FW} (chapter XI), we take only the $^{1}S_{0}$
scattering lengths and neglect the spin--triplet $^{3}S_{1}$ neutron--proton
contribution. This leads to an average 
$^{1}S_{0}$ scattering lengths, $a\simeq -20$ fm for symmetric matter while for neutron 
matter we simply take $a=-18.9$ fm. 

The $C$--term in the denominator would provide (in the Taylor
expansion) an additional higher--order contribution for the Lee-Yang
expression. We have however preferred to keep such term free and use the
coefficient as a parameter to adjust. We have then added explicitly other
terms in the functional, guided by Skyrme--type forces, to correctly
describe the EOSs at densities of interest in nuclear scales. The $%
\rho^{5/3} $ term mimics a term that would be produced (in a MF scheme) by a
velocity--dependent zero--range interaction. In particular, such term turns
out to be extremely important to improve the adjustment of the neutron
matter EOS in density ranges between the very dilute regime and the
saturation density. The $\rho^{\alpha+1}$ term in the EOS mimics a term that
would be generated by a density--dependent two--body zero--range interaction
like in Skyrme forces. 

We perform the adjustments, this time using as benchmark microscopic EOSs,
by including the constraints to describe the very low--density regime.
Benchmark data are: i) For neutron matter, the QMC AV4 results of Ref. \cite%
{geze} for values of $|ak_N|<$ 10 ($\rho <$ 0.05 fm$^{-3}$), and two
different sets of results for $|ak_N|>$ 10: the Friedman et al. results (FP)
of Ref. \cite{panda} or the Akmal et al. results (stiffer EOS) of Ref. \cite%
{akmal} [we call here the corresponding parameter sets YGLO (FP) and YGLO
(Akmal), respectively]; ii) For symmetric matter, the FP results and those
of Akmal et al. are very close from each others and we made a fit using only
the FP points. The values of the YGLO parameters are shown in Table III. 

\begin{table}[tbp]
\begin{tabular}{cccc}
\hline
& $C_{\beta}$ & $D_{\beta}$ & $F_{\beta}$ \\ 
& (fm$^2$) & (MeV.fm$^5$) & (MeV.fm$^{3+3\alpha}$) \\ \hline\hline
$\beta=0$ (FP) & 100.87 & -9264.18 & 9571.90 \\ 
$\beta=0$ (Akmal) & 70.19 & -8377.83 & 8743.85 \\ 
$\beta=1$ (FP) & 8.188 & -6624.87 & 6995.46 \\ \hline
\end{tabular}%
\caption{Values of the adjusted parameters obtained for the YGLO functional.
In all cases, $\protect\alpha=0.7$.}
\label{adju}
\end{table}

\begin{figure}[tbp]
\includegraphics[scale=0.6]{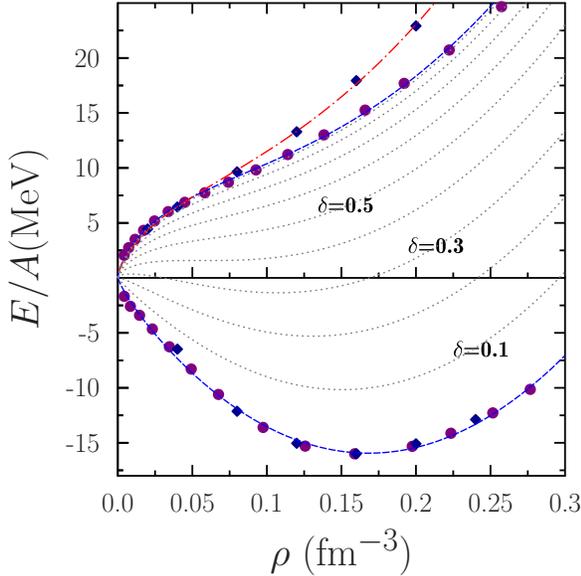}
\caption{EOSs from Akmal et al. \protect\cite{akmal} (blue diamonds) and
Friedman et al. (purple circles) in symmetric and neutron matter compared to
the YGLO (Akmal) (red dot-dashed curve) and YGLO (FP) (blue dashed curve)
results. The different grey dotted curves correspond to the YGLO(FP) EOSs
obtained for different asymmetry $\protect\delta$ from $0.1$ to $0.9$ by
steps of $0.1$ (see text).}
\label{NMSM}
\end{figure}

Figure \ref{neutronm} shows the results obtained with the YGLO functional
for neutron matter from the low--density regime to densities around
saturation. This evolution is compared with the SLy5 MF curve together with
several recent ab-initio calculations. We see that, with only four
adjustable parameters, the new functional gives results in agreement with
ab-initio calculations over the whole range of densities. In Fig. \ref{NMSM}%
, the EOSs obtained for symmetric and neutron matter are shown as a function
of $\rho$. In particular, one observes that the saturation properties are
well reproduced. The YGLO saturation density is $0.1683$ fm$^{-3}$ and
corresponds to an energy $E/A=-15.9$ MeV. The incompressibility modulus is
equal to 261.71 MeV. 

Starting from the two EOSs given above, one could use the standard parabolic
approximation to obtain the EOS in asymmetric matter. Introducing the
asymmetry parameter $\delta = (\rho_N-\rho_P)/(\rho_N+\rho_P)$ ($\rho_N$ and 
$\rho_P$ being the neutron and proton densities, respectively), the energy $%
E_\delta$ is given by 
\begin{eqnarray}
\frac{E_{\delta}}{A}(\rho) &=& \frac{E_{SM}}{A}(\rho) + S(\rho) \delta^2, 
\notag  \label{para}
\end{eqnarray}
where $S(\rho)$ is the symmetry energy, which can be computed, within the
parabolic approximation, as the difference of the EOSs of neutron and
symmetric matter. Corrections beyond such approximation are expected to be
small \cite{zuo1,zuo2}. As an illustration, several EOSs obtained with
different isospin asymmetries, from symmetric matter to pure neutron matter
are displayed in Fig. \ref{NMSM} by employing the YGLO(FP) functional for neutron 
matter.

\begin{figure}[tbp]
\includegraphics[scale=0.32]{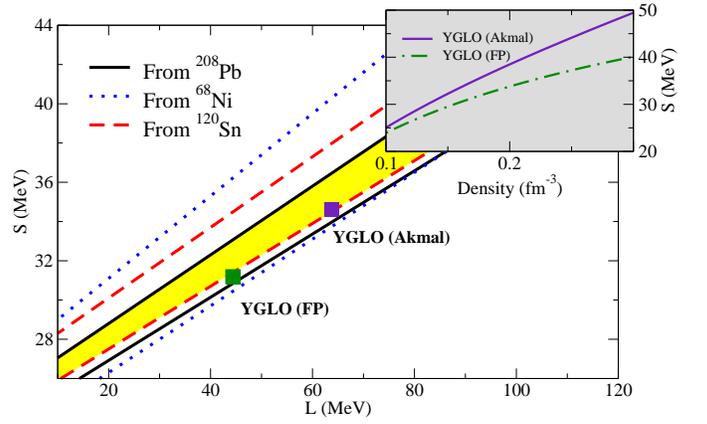}
\caption{Symmetry energy at saturation density as a function of its slope $L$%
. The black lines delimit the phenomenological area constrained by the
experimental determination of the electric dipole polarizability in $^{208}$%
Pb. The blue dotted lines delimit the area constrained by the same
measurement in $^{68}$Ni, and the red dashed lines refer to the measurement
done in $^{120}$Sn. The yellow area is the overlap. Inset: density dependence
of the Symmetry energy for the two YGLO parametrizations of this work. }
\label{sl}
\end{figure}

The density dependence of $S$ is known to be strongly connected to several
nuclear properties of astrophysical interest such as, for instance, the
proton fraction in neutron stars or to the thickness of neutron skins in
neutron--rich nuclei. In the inset of Fig. \ref{sl} the density dependences
of $S$ are illustrated for the two YGLO parameterizations. These density
dependences are comparable to those reported for instance in Ref. \cite%
{diepe}. As expected, the YGLO parametrization adjusted on the Akmal et al.
data provides a stiffer curve. A global quantity that characterizes the
symmetry energy evolution around saturation is its slope $%
L=3\rho_0(dS/d\rho)_{\rho=\rho_0}$. We present in Fig. \ref{sl} the points $%
(S,L)$ found in this work with respect to the phenomenological bands
provided in Fig. 5 of Ref. \cite{roca}. These bands are dictated by the
experimental determinations of the electric dipole polarizability in the
nuclei $^{208}$Pb, $^{68}$Ni, and $^{120}$Sn. The yellow area is the overlap
region of the three bands. We observe that both points are located at the
lower limit of the yellow area. Note that many phenomenological EDFs are
outside this band \cite{roca}.

In the present work, inspired by resummation techniques used in EFT, we
propose a local EDF that we call YGLO, able to describe the EOSs of
symmetric and neutron matter from very low densities to the saturation
density. We show that YGLO describes remarkably well saturation properties
of symmetric matter, including incompressibility, and leads to a density
dependence of the symmetry energy coherent with the phenomenological
indications provided by the measurement of the dipole polarizability in nuclei.


\begin{thebibliography}{99}
\bibitem{lee} T.D. Lee and C.N. Yang, Phys. Rev. 105, 1119 (1957).

\bibitem{hammer} H.W. Hammer and R.J. Furnstahl, Nucl. Phys. A 678, 277
(2000).

\bibitem{schafer} T. Sch\"afer, C.-W. Kao, and S.R. Cotanch, Nucl. Phys. A
762, 82 (2005).

\bibitem{skyrme} T.H.R. Skyrme, Philos. Mag. 1, 1043 (1956); Nucl. Phys. 9,
615 (1959).

\bibitem{vauth} D. Vautherin and D. M. Brink, Phys. Rev. C 5, 626 (1972).

\bibitem{gogny1} D. Gogny, Nucl. Phys. A 237, 399 (1975).

\bibitem{gogny2} J. Decharg\'e and D. Gogny, Phys. Rev. C 21, 1568 (1980).

\bibitem{chaba} E. Chabanat, P. Bonche, P. Haensel, J. Meyer, and R.
Schaeffer, Nucl. Phys. A 627, 710 (1997); 635, 231 (1998); 643, 441 (1998).

\bibitem{d1n} F. Chappert, N. Pillet, M. Girod, and J.-F. Berger, Phys. Rev. 
\textbf{C 91}, 034312 (2015).

\bibitem{Fur12} R.J. Furnstahl, \textit{Eft for DFT}. " Renormalization
Group and Effective Field Theory Approaches to Many-Body Systems. Springer
Berlin Heidelberg, 2012. 133-191.

\bibitem{FW} A.L. Fetter and J.D. Walecka, \textit{Quantum Theory of
Many--Particle Systems}, (McGraw-Hill, New York, 1971).

\bibitem{mog} K. Moghrabi, M. Grasso, G. Col\`o, and N.V. Giai, Phys. Rev.
Lett. 105, 262501 (2010).

\bibitem{new} C.J. Yang, M. Grasso, X. Roca-Maza, G. Col\`o, and K. Moghrabi, 
arXiv:1604.06278 [nucl-th].

\bibitem{gez} A. Gezerlis and G.F. Bertsch, Phys. Rev. Lett. 105, 212501
(2010).

\bibitem{bul1} A. Bulgac, Phys. Rev. A 76, 040502R (2007).

\bibitem{bul2} A. Bulgac, M.M. Forbes, and P. Magierski, The Unitary Fermi
Gas: From Monte Carlo to Density Functionals, Lecture Notes in Physics, Vol.
836 (Springer-Verlag, Berlin, Heidelberg, 2012), Chap. 9, pp. 305-373.

\bibitem{steele} J. V. Steele, arxiv:nucl-th/0010066v2

\bibitem{kaiser1} N. Kaiser, Nucl. Phys. A 860, 41 (2011).

\bibitem{geze} A. Gezerlis and J. Carlson, Phys. Rev. C 81, 025803 (2010).

\bibitem{wiringa} R. B. Wiringa and S. C. Pieper, Phys. Rev. Lett. 89,
182501 (2002).

\bibitem{Bis73} R. F. Bishop, Ann. Phys. \textbf{77}, 106 (1973).

\bibitem{panda} B. Friedman and V. Pandharipande, Nucl. Phys. A361, 502
(1981).

\bibitem{akmal} A. Akmal, V. R. Pandharipande, and D. G. Ravenhall Phys.
Rev. C 58, 1804 (1998).

\bibitem{zuo1} W. Zuo, I. Bombaci, and U. Lombardo, Phys. Rev. C 60, 024605
(1999).

\bibitem{zuo2} W. Zuo, et al., Eur. Phys. J. A 14, 469 (2002).

\bibitem{diepe} A.E.L. Dieperink, Y. Dewulf, D. Van Neck, M. Waroquier, and
V. Rodin, Phys. Rev. C 68, 064307 (2003).

\bibitem{roca} X. Roca-Maza, et al., Phys. Rev. C 92, 064304 (2015).

\bibitem{Gan08} S. Gandolfi, A. Yu. Illarionov, S. Fantoni, F. Pederiva, and
K. E. Schmidt, Phys. Rev. Lett. \textbf{101}, 132501 (2008).

\bibitem{Heb10} K. Hebeler and A. Schwenk Phys. Rev. \textbf{C 82}, 014314
(2010).
\end{thebibliography}
\end{document}